\documentclass[aps,physrev,twocolumn,superscriptaddress, amsmath,amssymb]{revtex4-2}
\usepackage{graphicx}
\usepackage{dcolumn}
\usepackage{bm}
\usepackage{siunitx}
\usepackage{xcolor}
\usepackage[dvipsnames]{xcolor}
\begin{document}

\title{Mass spectrometry of $^{75}$Zn ground and isomeric states from in-trap decay of $^{75}$Cu}

\author{M. Müller}
\email{Contact author: marius.mueller@mpi-hd.mpg.de}
\affiliation{Max-Planck-Institut für Kernphysik, 69117 Heidelberg, Germany}

\author{N. A. Althubiti}
\affiliation{School of Physics and Astronomy, The University of Manchester, Manchester M13 9PL, United Kingdom}
\affiliation{Physics Department, College of Science, Jouf University, Sakaka, Kingdom of Saudi Arabia}

\author{D. Atanasov}
\altaffiliation{Present address: Belgian Nuclear Research Centre, SCK CEN, Mol, Belgium}
\affiliation{Max-Planck-Institut für Kernphysik, 69117 Heidelberg, Germany}

\author{K. Blaum}
\affiliation{Max-Planck-Institut für Kernphysik, 69117 Heidelberg, Germany}

\author{R. B. Cakirli}
\affiliation{Max-Planck-Institut für Kernphysik, 69117 Heidelberg, Germany}

\author{T. E. Cocolios}
\affiliation{KU Leuven, Instituut voor Kern- en Stralingsfysica, 3001 Leuven, Belgium}

\author{F. Herfurth}
\affiliation{GSI Helmholtzzentrum für Schwerionenforschung GmbH, 64291 Darmstadt, Germany}

\author{S. Kreim}
\affiliation{Max-Planck-Institut für Kernphysik, 69117 Heidelberg, Germany}

\author{D. Lunney}
\altaffiliation{Present address:
CNRS-TRIUMF International Research Laboratory for Nuclear Physics, Nuclear Astrophysics and Accelerator Technology, Vancouver, Canada}
\affiliation{Université Paris-Saclay, CNRS/IN2P3, IJCLab, 91405 Orsay, France}

\author{V. Manea}
\altaffiliation{Present address: Université Paris-Saclay, CNRS/IN2P3, IJCLab, 91405 Orsay, France}
\affiliation{Max-Planck-Institut für Kernphysik, 69117 Heidelberg, Germany}
\affiliation{CERN, 1211 Geneva, Switzerland}

\author{N. Minkov}
\affiliation{Max-Planck-Institut für Kernphysik, 69117 Heidelberg, Germany}
\affiliation{Institute for Nuclear Research and Nuclear Energy,  Bulgarian Academy of Sciences,  BG-1784 Sofia, Bulgaria}

\author{D. Neidherr}
\affiliation{GSI Helmholtzzentrum für Schwerionenforschung GmbH, 64291 Darmstadt, Germany}

\author{M. Rosenbusch}
\altaffiliation{Present address: RIKEN Nishina Center for Accelerator-Based science, 351-0198 Wako, Saitama, Japan}
\affiliation{Universität Greifswald, Institut für Physik, 17487 Greifswald, Germany}

\author{L. Schweikhard}
\affiliation{Universität Greifswald, Institut für Physik, 17487 Greifswald, Germany}

\author{A. Welker}
\affiliation{CERN, 1211 Geneva, Switzerland}
\affiliation{Technische Universität Dresden, 01069 Dresden, Germany}

\author{F. Wienholtz}
\altaffiliation{Present address: Institut für Kernphysik, Technische Universität Darmstadt, 64289 Darmstadt, Germany}
\affiliation{CERN, 1211 Geneva, Switzerland}
\affiliation{Universität Greifswald, Institut für Physik, 17487 Greifswald, Germany}

\author{R. N. Wolf}
\altaffiliation{Present address: ARC Centre for Engineered Quantum Systems, School of Physics, The University of Sydney, Sydney, NSW 2006, Australia}
\affiliation{Max-Planck-Institut für Kernphysik, 69117 Heidelberg, Germany}

\date{\today}

\begin{abstract}
We report on high-precision mass measurements of the ground and first isomeric state of $^{75}$Zn, performed using the time-of-flight ion-cyclotron-resonance technique at the ISOLTRAP Penning-trap mass spectrometer at ISOLDE/CERN. The isomeric state was produced using in-trap decay of $^{75}$Cu. This marks the first direct investigation of the isomeric state of $^{75}$Zn via mass spectrometry. The isomer was observed at an excitation energy of \SI{123.7(20)}{keV}, in 2$\,\sigma$ agreement with the value previously determined through decay spectroscopy. In addition, our measurements correct a misassignment of the ground-state mass excess based on a previous measurement by Baruah \textit{et al.}, revising the value to \SI{-62681.0(21)}{keV}. To further investigate the earlier discrepancy, we explored the spin-parity assignments of the ground and isomeric states in $^{75}$Zn using Skyrme Hartree-Fock plus Bardeen-Cooper-Schrieffer theoretical calculations, given the absence of definitive experimental data. In light of the laser spectroscopy results from Wraith \textit{et al.}, our results add strong evidence for a spin-1/2 ground state, which would agree with large-scale shell-model predictions as well as explaining disagreements with the Monte Carlo Shell Model.
\end{abstract}

\maketitle

\section{Introduction}
The zinc isotopic chain is of particular relevance, being the first even-$Z$ chain above the closed $Z=28$ shell, nickel. As a result, the odd-$N$ zinc isotopes are sensitive to the evolution of neutron orbitals, which can be studied through the properties of their low-lying states.

 Both even-even and odd-$A$ Zn isotopes have been explored by various techniques, including $\gamma$-ray spectroscopy \cite{Ilyushkin2011, Cortes2018}, laser spectroscopy \cite{Wraith2017, Xie2019, Yang2018}, and mass spectrometry \cite{Baruah2008, Wolf2013_1}. Notably, \cite{Niikura2012} provided the first direct lifetime measurements of the 2$^+$ states in $^{72}$Zn and $^{74}$Zn, revealing evidence for a shape transition between $N=40$ and $N=42$ close to $Z=28$. These findings suggest that $^{72,74}$Zn exhibit $\gamma$-softness, while \cite{Hellgartner2023} further proposes that $^{72}$Zn has an average deformation close to maximal triaxiality in the $\gamma$ degree of freedom.

Additionally, laser spectroscopy studies have confirmed the existence of triaxial shapes in even-mass $^{72,74}$Zn isotopes and in the 5/2$^+$ isomeric state of $^{73}$Zn \cite{Yang2018}. Such triaxial deformations are consistent with those observed in the heavier even-even isotopes of Ge and Se within the same region \cite{Toh2013, Sun2014, Johnson1992, Guo2007, Yoshinaga2013, Bhat2014, Lettmann2017}.

In the field of mass spectrometry, the first measurement of $^{81}$Zn, along with the neutron-rich isotopes $^{71m,72-80}$Zn, was performed at ISOLTRAP in 2005 \cite{Baruah2008}. Following work extended the Zn mass measurements up to $^{82}$Zn \cite{Wolf2013_1}. In \cite{Baruah2008}, only a single long-lived state of $^{75}$Zn was observed, and its mass was assigned to the ground state. Subsequently, in 2011, Ilyushkin \textit{et al.} \cite{Ilyushkin2011} investigated the $\beta$-decay of $^{75}$Cu into states of $^{75}$Zn and, for the first time, identified an isomeric level at \SI{127.01(9)}{keV} that had not been observed in the mass-spectrometry study of \cite{Baruah2008}. To clarify this situation, we present here new experimental mass data for $^{75}$Zn obtained at ISOLTRAP \cite{Mukherjee2008, Lunney2017}. Because this state is not delivered in sufficient quantities directly from the target, we have produced it through in-trap $\beta$-decay \cite{Herlert2005, Ascher2019} of $^{75}$Cu in the ISOLTRAP preparation Penning trap. Alongside the experimental results, we also provide Skyrme Hartree-Fock plus Bardeen-Cooper-Schrieffer (HF-BCS) calculations for $^{75}$Zn. The details of the experimental setup, procedures, and data analysis are described in Sec.~II, while Sec.~III presents the experimental and theoretical results in the context of earlier work. Finally, Sec.~IV summarizes and concludes the study.
\section{Experiment and Data Analysis}
\subsection{Ion production and in-trap decay}
\begin{figure*}
\includegraphics[width=0.9\textwidth]{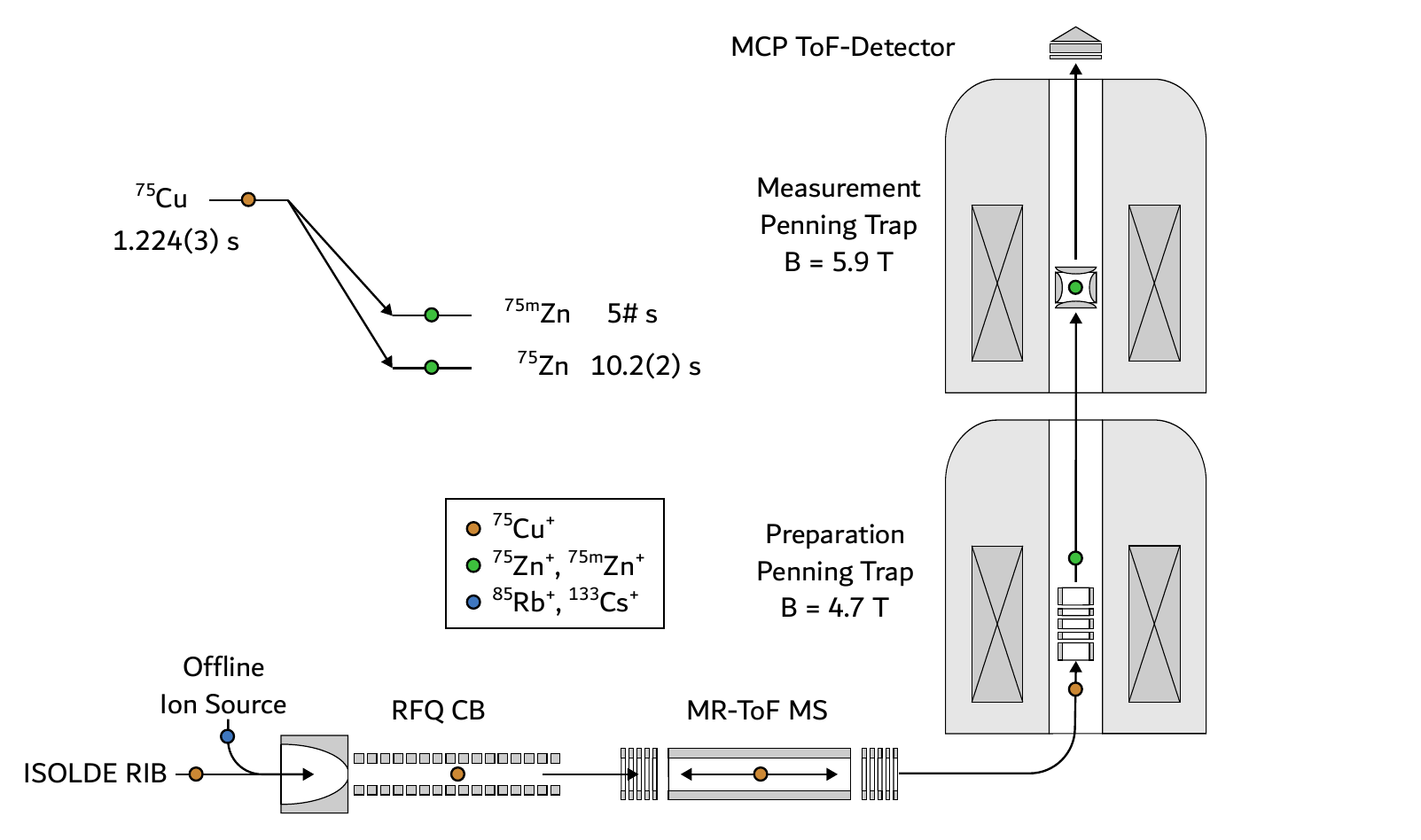}
\caption{\label{fig:setup} Schematic of the ISOLTRAP setup. The black arrows illustrate the ion path for the ions of interest, originating from the ISOLDE RIB source, and for the reference ions, produced by an offline alkali ion source, as they pass through the four ion traps of the ISOLTRAP system. The inset displays the decay scheme of $^{75}$Cu to the ground and isomeric states of $^{75}$Zn with the associated half-lifes \cite{NUBASE2020}.}
\end{figure*}
At the ISOLDE facility radioactive-ion beams (RIB) are produced via the isotope separation on-line (ISOL) method \cite{Köster2002, Catherall2017}. Unlike the experiment presented in \cite{Baruah2008}, $^{75}$Zn was not produced directly in target. Instead, the $\beta$-decay parent nucleus $^{75}$Cu was produced and transported to ISOLTRAP. To generate neutron-rich, medium-mass nuclei such as $^{75}$Cu, neutron-induced fission reactions were utilized, effectively suppressing the production of neutron-deficient contaminants. 

Proton bunches accelerated to a kinetic energy of \SI{1.4}{GeV} by CERN's proton synchrotron booster were directed onto a tungsten spallation target, also referred to as a \textit{neutron converter} \cite{Catherall2003}. The resulting spallation neutrons initiated fission reactions in a nearby thick uranium carbide target. The fission products diffused and effused from the target into a temperature-regulated transfer line, which transported the neutral atoms to an ion source. In this experiment, copper atoms were selectively ionized to the singly charged state using the resonance ionization laser ion source (RILIS) \cite{Fedosseev2012}.

In the following stage, the Cu$^+$ ions were accelerated using a \SI{30}{kV} potential and directed to the high-resolution separator (HRS) of ISOLDE. There, non-isobaric contaminants ($A \neq 75$) were removed from the ion beam using two sector magnets. The purified $^{75}$Cu$^+$ ions were subsequently delivered as a quasi-continuous beam to the ISOLTRAP setup (Fig.~\ref{fig:setup}), where time-of-flight ion-cyclotron-resonance (ToF-ICR) \cite{Koenig1995} measurements were carried out.

The ISOLTRAP measurement cycle begins in the helium-filled radio-frequency quadrupole cooler and buncher (RFQ CB) \cite{Herfurth2001}, which is located on a \SI{30}{kV} high-voltage platform, where incoming ions are decelerated, accumulated and cooled via collisions with helium buffer gas. After a cooling period of \SI{10}{ms}, the ions are extracted as a bunch, decelerated to approximately \SI{3.2}{keV} by a pulse-down drift-tube electrode, and then injected into a multi-reflection time-of-flight mass spectrometer/separator (MR-ToF MS) \cite{Wolf2013_2} for further purification. Inside the MR-ToF MS, another pulsed drift tube is used to further reduce the ion energy and to confine the ion bunch between two sets of mirror electrodes, forming a trapping potential. The ions oscillate back and forth for a predetermined number of revolutions. Because ions of different masses travel at slightly different velocities, the mass-dependent flight times result in a spatial separation of the ions of interest (IoI) from isobaric contaminants. This separation is then exploited to remove unwanted species during ejection using a Bradbury-Nielsen gate. The remaining $^{75}$Cu$^+$ ions are subsequently transferred to the preparation Penning trap, where they undergo helium buffer-gas cooling. 

Penning traps store charged particles via a superposition of a homogeneous magnetic field and an electric quadrupole field. This field configuration generates an ion motion, which can be decomposed into three eigenmodes: the axial mode, the cyclotron mode, and the magnetron mode. These eigenmodes have characteristic eigenfrequencies, the axial frequency $\nu_z$, the modified cyclotron frequency $\nu_+$, and the magnetron frequency $\nu_-$, which are described by the following equations in the absence of field imperfections:
\begin{equation}
    \nu_\mathrm{c} = \frac{1}{2\pi}\frac{q}{m} B\;,
\end{equation}
\begin{equation}
    \nu_\mathrm{z} = \frac{1}{2\pi}\sqrt{2 C_\mathrm{2} V_\mathrm{0} \frac{q}{m}}\;,
\end{equation}
\begin{equation}
    \nu_\mathrm{\pm} = \frac{\nu_\mathrm{c}}{2} \left[1\pm\sqrt{1-2\left(\frac{\nu_\mathrm{z} }{\nu_\mathrm{c}}\right)^2}\right]\;,
\end{equation}
with the cyclotron frequency $\nu_\mathrm{c}$, the electric charge $q$, the mass $m$, the magnetic flux density $B$, the quadratic expansion coefficient $C_\mathrm{2}$ of the electric potential, and the trapping voltage $V_\mathrm{0}$. 

An initial waiting time inside the preparation Penning trap serves to cool the axial and cyclotron mode of the stored ions, utilizing collisions with helium buffer gas. Typically, a cooling time of a few \SI{10}{ms} is sufficient to thermalize the modes. However, for $^{75}$Cu$^+$, an extended cooling time, which was varied between 1 and \SI{9}{s} was used, which allowed the majority of the ions to undergo in-trap $\beta$-decay, producing both the ground and first isomeric state of $^{75}$Zn$^+$. To remove any residual $^{75}$Cu$^+$ ions, \textit{mass-selective buffer gas cooling} was employed \cite{Savard1991}. 

For this, an initial \SI{25}{ms} quadrupole excitation at the estimated cyclotron frequency of $^{75}$Zn$^+$ was applied, which selectively couples the magnetron and cyclotron mode of this species via the \textit{sideband relation} $\nu_\mathrm{c}=\nu_\mathrm{-}+\nu_\mathrm{+}$. In combination with the collisional cooling of the cyclotron mode, this coupling effectively cools the magnetron mode, which is, in contrast to the other two modes, heated by the buffer gas collisions. Then a \SI{10}{ms} dipole excitation at the magnetron frequency, which is to first-order mass independent, was applied to drive all ions onto large magnetron orbits. The IoI were then selectively recentered by another \SI{25}{ms} quadrupole excitation at $\nu_\mathrm{c}$, after which a cooling time of \SI{25}{ms} served to rethermalize the cyclotron mode. 

During extraction from the preparation trap, an aperture blocked the contaminant ions, which remained on large magnetron orbits, allowing only the recentered IoI to pass through. Finally, the IoI were transferred to the measurement Penning trap, where their cyclotron frequency $\nu_c$ was determined via the ToF-ICR technique \cite{Koenig1995}.
\subsection{ToF-ICR mass measurement}
The ToF-ICR measurement begins with a \SI{10}{ms} phase-locked dipolar magnetron excitation \cite{Blaum2003}, which drives the ions onto large magnetron orbits. This is followed by a \SI{600}{ms} quadrupole radio-frequency excitation at $\nu_\mathrm{rf} \approx \nu_\mathrm{c}$. When on resonance, the excitation couples the magnetron and cyclotron mode, and increases the cyclotron radius in the process. Since $\nu_\mathrm{+} \gg \nu_\mathrm{-}$, this coupling substantially increases the magnetic moment. 

\begin{figure*}
\includegraphics[width=0.7\textwidth]{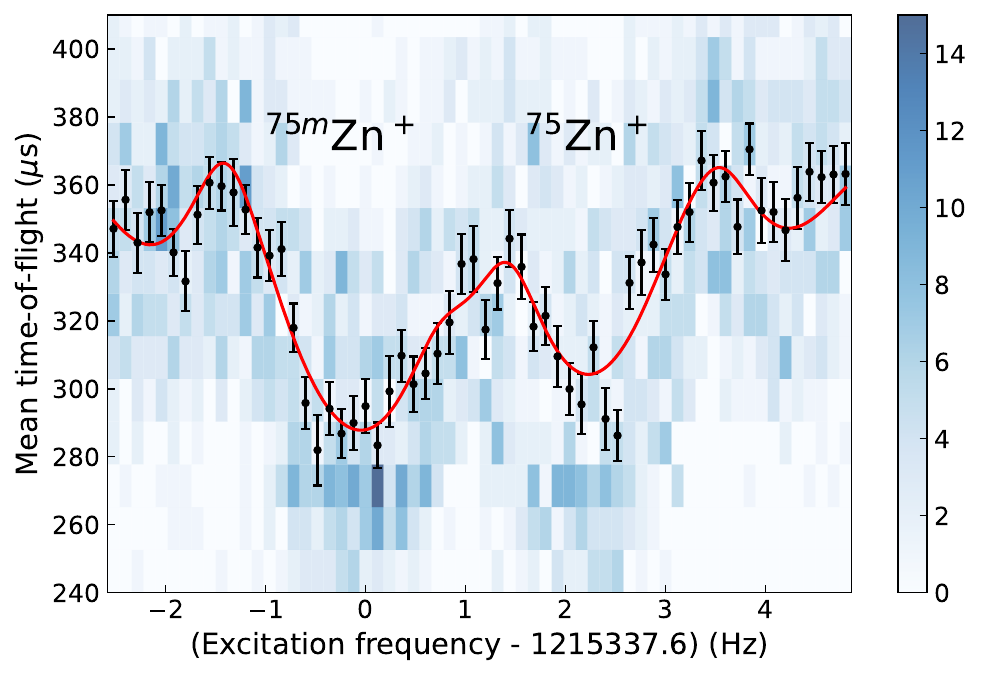}
\caption{\label{fig:resonance}One of the nine recorded ToF double-resonances with binned ion counts (blue), averaged ToF values (black), and a fit to the data using the model described in the text (red).}
\end{figure*}

After the excitation, the ions are ejected from the measurement trap towards a microchannel plate (MCP) detector. As they pass through the magnetic field gradient of the magnet's fringe field, the ions experience an acceleration proportional to their magnetic moment. By recording the ToF as a function of excitation frequency, ToF resonances are obtained. A minimum corresponds to resonant coupling at the cyclotron frequency $\nu_\mathrm{c}$. An example of such a ToF resonance, obtained in the $^{75}$Zn run, is shown in Fig.~\ref{fig:resonance}.

A total of nine ToF resonances were recorded over the course of two days, with a cumulative ion count of 10653. These measurements were periodically interrupted to acquire reference ToF spectra using $^{85}$Rb$^+$ ions from an offline alkali surface ion source. The reference measurements served to calibrate the magnetic field in the measurement trap and to determine the dimensionless mass~ratio
\begin{equation}
    r = \frac{m_{\mathrm{ioi}}}{m_{\mathrm{ref}}} = \frac{\nu_{\mathrm{c,ref}}}{\nu_{\mathrm{c,ioi}}} \;,
\end{equation}
where the charge ratio cancels out, as only singly charged ions were used, and the magnetic field dependence is eliminated through temporal interpolation. 
\subsection{Data Analysis}
The individual data sets were analyzed with the ToF-ICR software package EVA \cite{EVA}, which employs the line shape described in \cite{Koenig1995}. To correct for systematic shifts caused by ion-ion interactions, a so-called \textit{z-class analysis} was performed \cite{Kellerbauer2003_2}. In this procedure, the cycles of each ToF spectrum were grouped according to their count rate to generate sub-spectra. These sub-spectra were then fitted, and the z-class-corrected cyclotron frequency was obtained by linear extrapolation to a count rate equal to the detector efficiency ($\epsilon \approx 0.3$). This corresponds to a single ion per cycle, and therefore, no shift due to ion-ion interaction. 

Additional consistency checks were carried out by comparing the z-class analysis with fits to the unfiltered data as well as with fits restricted to cycles containing at most two ions. Except for one low-statistics spectrum, affected by a technical stop and therefore excluded from the final results, all analyses agreed within their uncertainties.

Systematic contributions were accounted for by adopting a relative mass-dependent shift of $1.6 \times 10^{-10}$ u$^{-1}$ and a residual statistical uncertainty of $8 \times 10^{-9}$, as established in \cite{Kellerbauer2003_2, Kellerbauer2003} from studies of carbon clusters at ISOLTRAP. Cross-check measurements of $^{85}$Rb$^+$ against $^{133}$Cs$^+$, performed before and after the zinc measurements, yielded frequency ratios consistent with the atomic mass evaluation (AME2020) \cite{AME2020}. Ultimately, the dominant contribution to the uncertainty in the mass determination remained statistical. For the final results presented below, only the z-class evaluation was adopted.
\section{Results and Discussions}
\begin{figure*}
\includegraphics[width=1.0\textwidth]{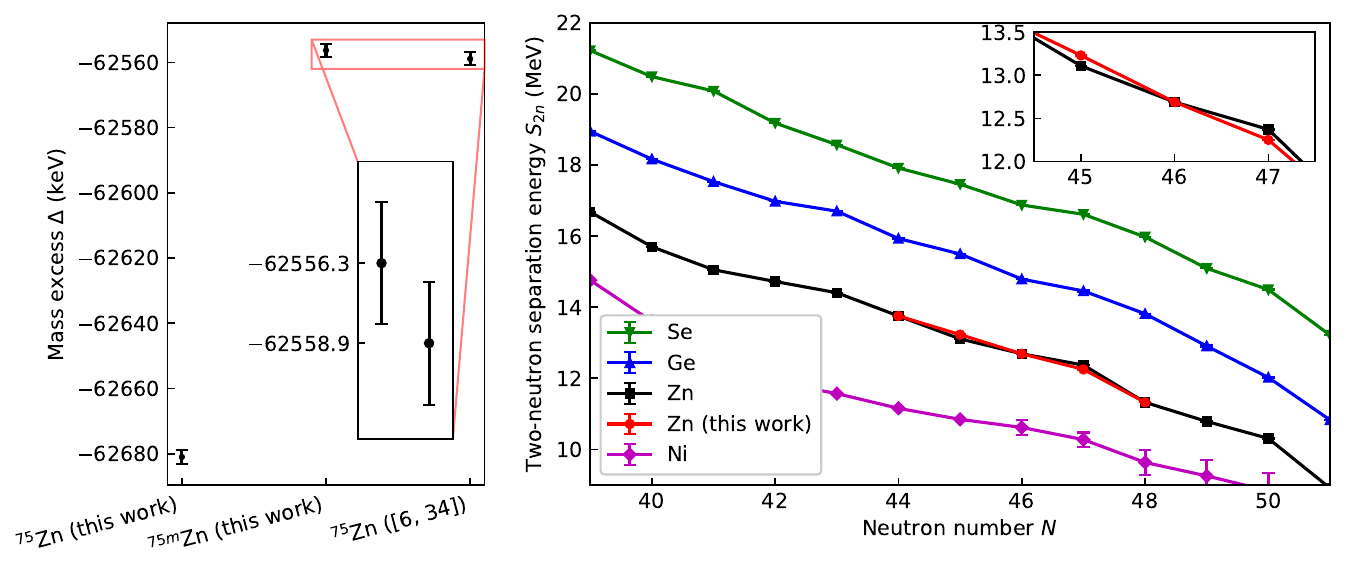}
\caption{\label{fig:results} (Left): Comparison of the mass excess values obtained in this work with those adopted in AME2020. The measured isomeric mass agrees within one combined standard deviation with the AME2020 value, determined \SI{100}{\%} by \cite{Baruah2008}, indicating a previous misassignment.  (Right): Two-neutron separation energies for nuclei in the vicinity of $^{75}$Zn. Incorporating the present experimental results yields a smoother $S_{2n}$ trend. The inset highlights the adjusted values at $N=45$ and $N=47$.}
\end{figure*}
\begin{table}[b]
\caption{\label{tab:resultsTable}%
Experimental results obtained in this work}
\begin{ruledtabular}
\begin{tabular}{cccc}
Ion & Reference & Mass ratio r & Mass Excess \\ 
 &  & &  $\Delta$ (keV) \\  \hline 
\rule{0pt}{1\normalbaselineskip}
$^{75}$Zn$^+$  & $^{85}$Rb$^+$ & $0.882\,476\,331(26)_{stat}(7)_{sys} $     & -62681.0(21)           \\
$^{75m}$Zn$^+$ & $^{85}$Rb$^+$ & $0.882\,477\,908(24)_{stat}(7)_{sys} $      & -62556.3(20)          
\end{tabular}
\end{ruledtabular}
\end{table}
The results of the data analysis are summarized in Table~\ref{tab:resultsTable} and displayed in the left panel of Fig.~\ref{fig:results}, together with the AME2020 mass value, which is based on the measurement reported in \cite{Baruah2008}. The comparison of the mass excesses reveals a clear correspondence between the AME2020 value and the isomeric-state mass excess obtained in this work. The value from \cite{Baruah2008} (\SI{-62558.9+-2.0}{keV}) is in \SI{1}{\sigma} agreement with the new isomer mass. Our result is also in agreement with a recent MR-ToF measurement from RIKEN \cite{Xian2025} (\SI{-62571+-13}{keV}), who also assumed they had measured the ground state, but our measurement is over six times more precise.

From the ToF-ICR double-resonance measurements of the two $^{75}$Zn states following in-trap decay of $^{75}$Cu, we could determine the branching ratio between the two states. In the in-trap decay measurements, the isomeric ratio was $N_m/(N_g+N_m) = 0.60(7)$ for the shortest storage time of \SI{1}{s}, neglecting shifts in the ratio due to decays to $^{75}$Ga. For the longest storage time of \SI{9}{s}, the ratio shifted to $0.74(3)$, indicating that the half-life of the isomeric state exceeds the \SI{5}{s}, estimated from trends in neighboring nuclei \cite{NUBASE2020}, and is likely larger than the half-life of the ground state. However, limited statistics preclude a definitive determination of the isomeric half-life. 

The mass of $^{75}$Zn produced directly at ISOLDE was first measured in 2005 by Baruah \textit{et al.} \cite{Baruah2008}, where a single ion species was observed and attributed to the ground state of $^{75}$Zn. Later, Ilyushkin \textit{et al.} \cite{Ilyushkin2011} reported the level scheme of $^{75}$Zn based on the $\beta$-decay of $^{75}$Cu. Their $\gamma$-ray spectroscopy results tentatively assigned the ground and isomeric state of $^{75}$Zn spin-parities of 7/2$^+$ and 1/2$^-$, respectively.

It is well established that laser spectroscopy provides information on nuclear spin, but it cannot unambiguously determine on its own which state corresponds to the ground state and which to an isomeric state. Laser spectroscopy studies reported in \cite{Wraith2017, Xie2019} identified levels with spins 7/2 and 1/2. The corresponding parity assignments were obtained from nuclear magnetic moment measurements. The laser-spectroscopy results are consistent with the spin values proposed by the $\beta$-decay results discussed earlier \cite{Ilyushkin2011}, but cannot by themselves decide on the ordering between the two states.

An additional piece of information with respect to spin assignment comes from the production ratio of the two states. In laser-spectroscopy studies performed at ISOLDE, it was established that the production ratio between the 7/2 and 1/2 state in $^{75}$Zn is 6:1 \cite{Wraith2017}. The corresponding experiment used a similar target-ion-source assembly and production mechanism (neutron-induced fission of uranium) as the ISOLTRAP experiment of \cite{Baruah2008}. The determined ratio is consistent with the expected yield of approximately \SI{85}{\%}, assuming the Madland-England model \cite{Madland1977, Sears2021}. This suggests that the mass measured in the experiment of Baruah et al. \cite{Baruah2008} is the 7/2 state. Since this mass value corresponds to the larger of the two masses measured in this work, it appears that the 7/2 state should be the excited state, while the ground state in $^{75}$Zn should have spin 1/2.

If the 7/2$^+$ level is not the ground state, it implies that the ground state is 1/2$^-$. In such a case, the spin–parity assignments of several excited states would need to be reconsidered due to the established $\gamma$-ray transitions connecting them. For instance, the \SI{475}{keV}  level, previously assigned as 9/2$^+$ \cite{Ilyushkin2011}, would need to be reinterpreted as 1/2$^{+,-}$, 3/2$^{+,-}$ or 5/2$^{+,-}$.

The right panel of Fig.~\ref{fig:results} shows the two-neutron separation energy 
\begin{equation}
\begin{aligned}
    S_{2n}(Z, N) =& BE(Z, N)-BE(Z,N-2) \\
     =& \left[m(Z,N-2)+ 2m_n-m(Z,N)\right]c^2 \;,
\end{aligned}
\end{equation}
with the binding energy $BE$, the neutron mass $m_n$ and the speed of light $c$,
as a function of neutron number for Ni, Zn, Ge, and Se isotopes. Since this work distinguishes between the ground- and isomeric-state masses of $^{75}$Zn, new $S_{2n}$ values at $N=45$ and $N=47$ have been obtained, as highlighted in the inset of the figure. These results make the downward linear trend in $S_{2n}$ with increasing neutron number smoother.

According to theoretical studies on the even-$Z$, $N=45$ isotones \cite{Ilyushkin2011}, the 7/2$^+$ state arises from Coriolis mixing among three neutrons occupying the 1g$_{9/2}$ orbital, while the 1/2$^-$ state is a consequence of the nearby 2p$_{1/2}$ orbital. Low-lying 7/2$^+$ states in this mass region have long been interpreted as dominated by a (g$_{9/2})^3$ configuration \cite{Mihelich1951}. 

Theoretical predictions for the ground and low-lying excited states of $^{75}$Zn vary considerably across the literature \cite{Wraith2017, Xie2019}. In \cite{Wraith2017}, large-scale shell-model calculations were performed using different model spaces and effective interactions. The simplest configuration employs a $^{56}$Ni core with the JUN45 effective interaction \cite{Honma2009}. Beyond this, an extended model space was explored that includes additional proton and neutron excitations. For this extended model, two interactions were applied: the modified A3DA interaction (A3DA-m) within the Monte Carlo Shell Model (MCSM), and a modified LNPS interaction (LNPS-m), for the description of odd-A zinc isotopes in the range $^{71-79}$Zn \cite{Tsunoda2014, Otsuka2001, Lenzi2010}.

The resulting predictions for $^{75}$Zn differ significantly depending on the chosen interaction: the ground-state spin-parity is 5/2$^+$ with A3DA-m, 1/2$^-$ with LNPS-m, and 9/2$^+$ with JUN45. Notably, none of these theoretical results agree with one another, nor do they match the experimentally proposed 7/2$^+$ ground state reported in \cite{Ilyushkin2011}.

In addition, we present our own calculations based on the HF-BCS approach developed in \cite{Bonneau2015}. Within this framework, the spin and parity of the ground and excited states of an odd-mass nucleus are determined through a self-consistent blocking of the single-particle (s.p.) orbitals located near the Fermi level in the s.p. spectrum of the corresponding even-even core nucleus. The ground state is identified as the configuration with the lowest total energy obtained from the self-consistent solution, while the excited states correspond to quasi-particle (q.p.) states with successively higher total energies. The BCS pairing constants are initially adopted and subsequently adjusted to reproduce the energy of the first excited $2^{+}_{1}$ state of the even-even core nucleus, as determined from the microscopically calculated moment of inertia (for further details, see \cite{Minkov2024}).

We applied this procedure to $^{75}$Zn using the well-established SIII Skyrme interaction \cite{Beiner1975} as a starting point, followed by several additional interactions, including SLyIII.0.8, SkM*, and SLy4, among others. Across all tested interactions, our calculations consistently predict the ground-state spin and parity of $^{75}$Zn to be 5/2$^+$. In addition, the calculations indicate the presence of low-lying excited states with spin and parity 7/2$^{+}$, 1/2$^{-}$, and 9/2$^{+}$, corresponding to one-q.p. excitations whose ordering depends on the adopted pairing strengths. Overall, these results are in agreement with the MCSM predictions using the A3DA-m interaction reported in \cite{Wraith2017}.

We further applied the same calculation procedure to the neighboring odd-mass isotopes $^{73,77}$Zn. For $^{73}$Zn, the results indicate that the ground-state spin depends on the adopted pairing strengths: either 1/2$^{-}$, with 5/2$^+$ appearing as a one-q.p. excitation, consistent with the MCSM calculations using the A3DA-m interaction and the experimental data \cite{Yang2018, Wraith2017}, or 5/2$^+$, as predicted by the MCSM calculations with the LNPS-m interaction \cite{Wraith2017}.
For $^{77}$Zn, our calculations consistently predict a 7/2$^{+}$ ground state, in agreement with both A3DA-m and LNPS-m MCSM calculations as well as the reported experimental data \cite{Wraith2017}. In this case, the 9/2$^{+}$ and 5/2$^+$ configurations appear as one-q.p. excited states.
\begin{table}[]
\caption{\label{tab:resultsTable2}%
Theoretical ground state spin-parities of $^{75}$Zn.}
\begin{ruledtabular}
\begin{tabular}{cccc}
Nucleus & Interaction & J$^{\pi}$ & Reference  \\ \hline
\rule{0pt}{1\normalbaselineskip}
$^{75}$Zn & SIII, SLyIII.0.8, SkM*, SLy4 & 5/2$^+$ & this study  \\ 
           & A3DA-m & 5/2$^+$ & \cite{Wraith2017} \\
           & LNPS-m & 1/2$^-$ & \cite{Wraith2017} \\ 
           & JUN45 & 9/2$^+$ & \cite{Wraith2017} \\
\end{tabular}
\end{ruledtabular}
\end{table}
Summarizing the results of the theoretical investigations discussed above, it becomes evident that for the neighboring isotopes $^{73}$Zn and $^{77}$Zn, both the HF-BCS and MCSM approaches are able, under certain conditions such as appropriate interaction choices or pairing strengths, to reproduce the experimentally established ground-state spin and parity. However, for $^{75}$Zn, none of the theoretical frameworks predicts the 7/2$^{+}$ state as ground state.

As our discussion focuses primarily on the spin–parity assignments of the ground and isomeric states, the detailed level scheme is not presented here. Given the inconsistencies among the theoretical models, only the calculated ground-state spin–parity values are summarized in Table~\ref{tab:resultsTable2}.

We presented experimental evidence that the ground state of $^{75}$Zn has spin 1/2. However, this relies on the assumption that in both the previous mass spectrometry \cite{Baruah2008} and laser spectroscopy \cite{Wraith2017} experiments, the production conditions were identical. While both experiments used neutron-induced fission of uranium at the same primary-beam energy, one cannot exclude differences in the position of proton impact on the neutron converter or the fraction of protons scattered on the target, determining proton-induced fission. While it is unlikely that these factors alone could lead to a complete reversal of the $^{75}$Zn state predominantly produced from the fission of uranium, the evidence presented in this work for a 1/2$^-$ ground state in $^{75}$Zn should be regarded as indirect. To directly confirm the assignments for the ground and isomeric states, further detailed spectroscopic investigations are required. 

\section{Conclusions}
In this work, the masses of both the ground and isomeric state of the $^{75}$Zn nucleus were measured at ISOLTRAP. The isomeric state was identified for the first time by mass spectrometry at an excitation energy of \SI{123.7(20)}{keV}. In the context of previous experimental observations \cite{Baruah2008}, questions about the spin–parity assignments of the ground and isomeric states of $^{75}$Zn arise. Numerous theoretical studies have addressed this nucleus, yet their predictions are often inconsistent with one another. The theoretical model presented in this work also yields results that differ from some of those reported previously. 

Based on the results of earlier ISOLDE experiments \cite{Baruah2008, Wraith2017}, our findings provide evidence supporting a 1/2$^-$ ground state in $^{75}$Zn. Nonetheless, additional experimental investigations are required to conclusively establish this assignment. Such studies would provide valuable benchmarks for refining and improving theoretical nuclear-structure models.

\begin{acknowledgments}
We thank R.F.~Casten, U.~Köster and K.~Chrysalidis for the discussions. We thank the ISOLDE technical group and the ISOLDE Collaboration for support. We also acknowledge support by the BMBF (05P12HGCI1 and 05P15ODCIA), the FWO-Vlaanderen (I002619N and I001323N), the Max Planck Society, French IN2P3, the Alliance Program of the Helmholtz Association Contract No. HA216/EMMI, and the STFC under Grants No. ST/L005743/1 and No. ST/L005816/1. N. M. acknowledges support by the Bulgarian National Science Fund (BNSF) under Contract No. KP-06-N98/2. A. W. acknowledges support by a Wolfgang Gentner Scholarship.
\end{acknowledgments}

\end{document}